\documentclass[10pt,aps,prl,twocolumn,showkeys]{revtex4-1}
\usepackage{amsmath}
\usepackage{graphicx}
\usepackage{epstopdf}
\usepackage{threeparttable}
\usepackage{color}

\usepackage{dcolumn}
\usepackage{bm}
\usepackage{amssymb}
\usepackage[mediumspace,mediumqspace,Grey,squaren]{SIunits}

\begin{document}
\title{Screen-printed and spray coated graphene-based RFID transponders}
\author{K. Jaakkola$^1$, V. Ermolov$^1$, P. G. Karagiannidis$^2$, S. A. Hodge$^2$, L. Lombardi$^2$, X. Zhang$^2$, R. Grenman$^1$, H. Sandberg$^1$, A. Lombardo$^2$, A. C. Ferrari$^2$}
\affiliation{$^1$VTT Technical Research Centre of Finland, Espoo 02044, Finland}
\affiliation{$^2$Cambridge Graphene Centre, 9 JJ Thomson Avenue, University of Cambridge, Cambridge, CB3 0FA, UK}

\begin{abstract}
We report Ultra-High-Frequency (UHF, 800MHz-1GHz) Radio Frequency Identification (RFID) transponders consisting of printed dipole antennas combined with RFID microchips. These are fabricated on Kapton via screen printing and on paper via spray coating, using inks obtained via microfluidization of graphite. We introduce a hybrid antenna structure, where an Al loop (small compared to the overall size of the antenna) is connected to a microchip with the double function of matching the impedances of antenna and microchip and avoiding bonding between exfoliated graphite and chip. The transponders have reading distance$\sim$11m at UHF RFID frequencies, larger than previously reported for graphene-based RFID and comparable with commercial transponders based on metallic antennas.
\end{abstract}

\maketitle

\section{Introduction}
Radio frequency identification (RFID) is a ubiquitous technology\cite{FinkBook}, with applications in access control\cite{FinkBook}, contactless payment\cite{LacmMIPRO}, electronic passports\cite{FinkBook}, supply chain management\cite{MusaGJFSM17}, healthcare\cite{WambIJIM33}, food packaging\cite{BijiJFS52} and animal identification\cite{GeerBook}. It is also the cornerstone of the so-called "Internet of Things" (IoT)\cite{AshtRFIDJ22}, where people and devices are seamlessly integrated in a decentralized common set of resources, creating a convergence of the physical realm with human-made virtual environments\cite{BuyyBook}. Within IoT, every "thing" is connected\cite{BuyyBook}, and the already widespread RFID technology is likely to become even more ubiquitous, combining additional functionalities such as sensing\cite{NairRFID-TA2012,ManzRFID-TA2012} and energy harvesting\cite{ParkRFID-TA2014,FerdRSER58}.

The basic elements of a typical RFID system are tags\cite{FinkBook} and readers\cite{FinkBook}, which exchange information via radio waves\cite{FinkBook}. Tags comprise integrated circuits containing a memory to store the tag identity (ID) and the reading/writing circuitry\cite{FinkBook}. Tags communicate with the reader via a suitable antenna, which typically has the double role of drawing energy from the reader to energize the integrated circuit\cite{FinkBook}, and exchange data with the reader\cite{FinkBook}. RFID offers advantages over other identification technologies, such as barcodes\cite{FinkBook}, since an RF tag does not need to be in sight of the reader and can, therefore, be embedded in objects\cite{FinkBook}. Also, RFID allows simultaneous reading of several tags\cite{FinkBook}, making the identification process very fast, typically a few milliseconds for passive (i.e. powered by the reader through the antenna) RFID tags\cite{FinkBook} and even shorter for active ones (i.e. powered by a battery)\cite{FinkBook}.

RFID tags should combine mechanical robustness (e.g. to tolerate vibrations)\cite{FinkBook}, light weight (typically$<$10g) \cite{FinkBook}, compact dimensions ($\sim$cm)\cite{FinkBook}, reliability\cite{FinkBook} and low cost($<$0.05\$)\cite{SarmaWP}. Mechanical flexibility might also be required (especially for IoT\cite{KaleBook}), adding specific challenges not present on rigid systems, such as shifts in resonant frequency\cite{KaleBook}, and return loss (i.e. reflected power loss caused by antenna input impedance mismatch)\cite{KaleBook} and changes in effective capacitance (i.e. the ratio of the change in charge to the corresponding change in potential)\cite{KaleBook}, radiation pattern distortion\cite{KaleBook} and gain degradation\cite{KaleBook}. Different operational scenarios also introduce additional complexity, e.g. proximity to tissues in wearable devices\cite{KaleBook}.

Large volume (several millions of units)\cite{SarmaWP} and low cost ($<$0.05\$ per unit)\cite{SarmaWP} manufacturability is essential, as it is expected that over one trillion IoT devices will be deployed by 2025\cite{McKi2015,RNRM18}. The most common RFID tags, consisting of a planar electric dipole antenna\cite{Rao2005,Marrocco2008,Finkenzeller2010}, are fabricated from a metallized plastic foil by acidic etching\cite{FinkBook}. However, this process results in metal waste\cite{Cui2016}, which is also environmentally harmful\cite{Cui2016}.

Printing is a promising alternative\cite{KaleBook}, as it combines high volume production (e.g. an industrial screen printer can print areas$>3m\times$6m in a single pass\cite{kippax}) and, at the same time, avoids chemical etching and material wastage. Ag inks are typically used for printed RFID\cite{NikiIEEEAPS2B, PongISAP12}, since they have high conductivity$\sim10^6$S/m\cite{DearMRC26}. However, the Ag cost is very high ($\sim$800-1000\$/kg)\cite{LouwSEMS}. Printed Ag films have limited flexibility, breaking$\sim75\%$ strain\cite{Merilampi2009} and resistance increase up to$\sim15\%$ upon bending\cite{Merilampi2009}. Moreover, they can be toxic and carcinogenic\cite{Sondergaard2014}.

Printed graphene layers can be an alternative to printed metals\cite{FerrNS7} as graphene combines good conductivity\cite{FerrNS7} and mechanical robustness\cite{FerrNS7}. Graphene can be dispersed in solvents (such as NMP\cite{BonaMT15} or water\cite{BonaMT15}), doped\cite{BonaMT15} or functionalized\cite{BonaMT15}. The surface resistivity of single layer graphene (SLG) at radio (300KHz to 300MHz) and microwave (300MHz to 300GHz) frequencies is higher than metals\cite{PeruLAP2012}, resulting in losses\cite{PeruLAP2012} that prevent its use in antennas with high ($>90\%$) efficiency (i.e. ratio between power irradiated by the antenna and power supplied)\cite{PeruLAP2012}. The SLG conductivity can be tuned by field effect\cite{NovoS306}. However, the changes are mostly in the real part\cite{PeruLAP2012}, while in the imaginary part these are small up to$\sim$100GHz\cite{HornPRB83,Awan2DM3}, resulting in limited reconfigurability (i.e. tunability of radiation frequency, pattern or polarization)\cite{PeruEMC2013}.

Thick ($>1\mu$m) exfoliated graphite films, consisting of few-layer graphene (FLG) flakes, can overcome such limitations, showing sheet resistances R$_S<2\Omega/ \square$)\cite{Karagiannidis2017}, corresponding to conductivities$>$10$^4$S/m\cite{Karagiannidis2017}. These can also be deposited over large ($m^2$) areas by screen printing or spray coating.

Screen printing is a common industrial technique for roll-to-roll patterned deposition\cite{kippax}. Typical formulations of screen inks contain a conductive filler, such as Ag particles\cite{Merilampi2009}, and insulating additives (e.g. stabilizers and binders)\cite{Birkenshaw}, at a total concentration$>$100g/L\cite{Birkenshaw}. Of this,$>$60g/L consists of the conductive filler needed to achieve sufficiently high ($>10^6$S/m) conductivities\cite{Merilampi2009, Hyun2015}. Spray coating is also suitable for roll-to-roll production\cite{Mori}. To the best of our knowledge, there are no reports on spray coated graphene-based antennas. However, spray coated FLG films with similar specifications to those needed for RFID antennas (R$_S\sim$6$\Omega/\square$ and thickness$\sim$8$\mu$m) were reported for use in Electromagnetic Interference (EMI) shielding\cite{AcquISEC}.

A number of antennas based on solution-processed FLG films have been reported\cite{Huang2015,Akbari2016,Arapov2016,Kopyt2016,LammIEEEAWPL99,PamNC9}. Their reduced performance in gain and radiation efficiency compared to metallic antennas (typically over one order of magnitude\cite{PeruEMC2013}) is compensated by other functionalities, such as mechanical flexibility\cite{Arapov2016}. RFID transponders, based on FLG film antennas combined with RFID integrated circuits, were demonstrated\cite{Akbari2016, Arapov2016, Kopyt2016}, showing typical reading distance up to$\sim$9m\cite{PamNC9}. This is smaller than commercial RFIDs, providing$>$10m\cite{Akbari2016, Arapov2016, Kopyt2016}.

The input impedance of a typical RFID microchip at operating frequencies (865-868 MHz in Europe and 915MHz in US\cite{FinkBook}) is capacitive\cite{FinkBook,Impinj}, with a real part lower than the absolute value of the reactance\cite{FinkBook,Impinj}. Thus, to match the impedance conjugately, i.e. to ensure that both microchip and antenna are electrically compatible with each other, the impedance of the antenna should be the complex conjugate to that of the microchip at the frequency of operation\cite{NikiIEEEAPS2B}. A two-branch dipole antenna might not have such a point on its impedance curve because of design\cite{PozaBOOK}, dimensions\cite{PozaBOOK} or materials used\cite{PozaBOOK}. The conjugate impedance match between microchip and antenna can be achieved by forming a loop inductor parallel to the feeding point on the antenna conductor\cite{Finkenzeller2010}.

Here, we report RFID transponders consisting of graphitic antennas either screen printed on Kapton or sprayed on paper, coupled with RFID chip through Al inductive loops, ensuring impedance matching, i.e. that the impedance of the antenna is the complex conjugate impedance of the microchip at the frequency of operation. The Al loop is significantly smaller than the overall antennas size, therefore minimizing use of metal and not compromising the flexibility of the overall transponder. These have reading distances up to$\sim$11m in the relevant UHF RFID bands: 865.6-867.6MHz (Europe) and 902-928 MHz (USA and Japan), larger than graphene-based RFID tag previously reported\cite{Akbari2016, Arapov2016, Kopyt2016,PamNC9} and comparable with commercial RFID transponders\cite{Datasheet}.
\section{Results and Discussion}
\begin{figure}
\centerline{\includegraphics[width=90mm]{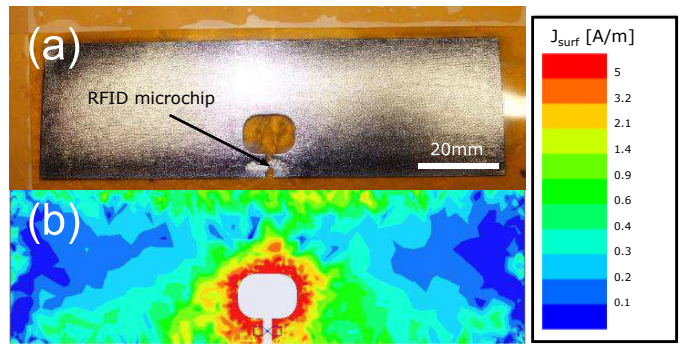}}
\caption{a)Antenna with FLG inductor. b) Simulated current distribution. J$_{surf}$ is the surface current density in A/m}
\label{Figure1}
\end{figure}
\begin{figure}
\centerline{\includegraphics[width=90mm]{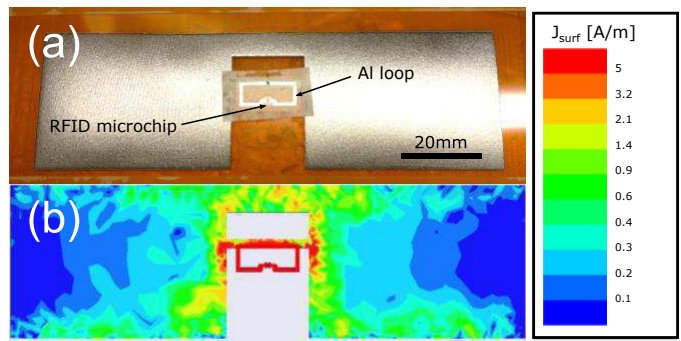}}
\caption{(a) Hybrid antenna with metal inductor. The larger structure is the printed FLG, while the inner loop is the Al inductor. b) Simulated current distribution. J$_{surf}$ is the surface current density in A/m.}
\label{Figure2}
\end{figure}

The antennas are designed using the electromagnetic simulation software High Frequency Structure Simulator (HFSS) 15 (Ansys Inc. USA), assuming R$_S\sim3\Omega/ \square$, as typical for dried FLG films produced by microfluidizaton\cite{Karagiannidis2017}. The two main parameters of a transponder antenna are the input impedance\cite{FinkBook}, to match the antenna with the transponder microchip, and the radiation efficiency, defined as the ratio of power radiated by the antenna and power supplied\cite{PozaBOOK}.

We use a Impinj Monza R6 UHF RFID microchip, with a 96 bits memory. This employs unregulated codes and is compatible with a wide range of tag form factors\cite{Impinj}. The input impedance is 16-j139$\Omega$ at 915MHz\cite{Impinj}. This is prevalently capacitive, with a real part lower than the absolute value of the reactance. Thus, to match the impedances conjugately the antenna should have an impedance Z$_{ant}$=16+j139$\Omega$ at the same frequency, i.e. the antenna should be sufficiently inductive with a low real part of the impedance. In order to achieve this, a parallel inductor in the dipole antenna is implemented as a opening on the conductor\cite{Akbari2016,Arapov2016}.

We also introduce a hybrid structure in which we combine the printed FLG antenna with an Al inductive loop for impedance matching. The Al loop is significantly smaller than the overall size of the transponder, therefore minimizes the use of metals and does not compromise flexibility. The loop forms inductive coupling between microchip and the antenna FLG conductor. Thus, no direct connection of microchip to FLG film is required.

We design and simulate FLG antennas using both FLG inductors and Al inductive loops. Both designs are made for the same FLG R$_S\sim3\Omega/ \square$. The optimized outer dimensions of the antenna to work at 915 MHz with the FLG inductive loop, shown in Fig. \ref{Figure1}, are 114mm$\times$34mm and the dimensions of the opening are 13.3mm$\times$10.1mm. The outer dimensions of the hybrid antenna, Fig.\ref{Figure2}, are the same. The dimensions of the upper opening of the antenna are 18.3mm$\times$6mm, and those of the lower opening are 18.3mm$\times$20mm.

The main tunable parameters of the antennas, optimized by simulations, are the circumference of the loop and the length of the antenna. The first determines the input reactance of the antenna\cite{Akbari2016,Arapov2016}, while the latter determines the radiation resistance, i.e. the resistance caused by the radiation of electromagnetic waves from the antenna\cite{MiliBOOK}. In the hybrid antenna, a rectangular opening is added, rather than a loop, in order to minimize Eddy currents induced by the inductive loop, since these would increase losses and decrease radiation efficiency. Shape and dimensions of the opening are chosen to minimize Eddy currents without significantly affecting antenna conductivity. The inductive loop, with 14mm$\times$6mm outer dimensions, is made of 0.8mm wide and 9$\mu$m thick Al and is shown in Fig.\ref{Figure2}.

Table I summarizes the simulated parameters at 915MHz: input impedance Z$_{ant}$, attenuation due to impedance mismatch L$_{Z}$, radiation efficiency $\eta$, directivity (i.e. the ration between the maximum radiation intensity in the main beam and the average radiation intensity over all space) D$_{tag}$ and calculated read range, i.e the calculated maximum distance that the RFID tag can be read, R$_{read}$. The attenuation due to the impedance mismatch between antenna and microchip is be calculated from the impedances as\cite{PozaBOOK}:
\begin{equation}
L_{Z}=1-|(Z_{ant}-Z_{IC}^*)/(Z_{ant}+Z_{IC})|^2
\end{equation}
where Z$_{IC}$ is the complex impedance of the microchip. The forward-link (i.e. the transmission from the reader to the tag\cite{Niki_online}) read range can be calculated as\cite{Niki_online,Su}:
\begin{equation}
R_{read} = (c/4 \pi f)\times(P_{tx EIRP}D_{tag}\eta_{tag}L_{Z}/P_{IC sens})^{1/2}
\end{equation}
where c is the speed of light, f is the frequency, P$_{tx EIRP}$ is the equivalent isotropically radiated power (i.e. the measured radiated power in a single direction) of the reader device and P$_{IC sens}$ is the read sensitivity of the microchip (i.e. the minimum power required to activate the chip). P$_{tx EIRP}$=3.28W is the maximum allowed radiated power of a UHF RFID reader as defined by the European regulatory environment for radio equipment and spectrum\cite{ECC}. P$_{IC sens}$=-20dBm, as specified for the Impinj Monza R6 microchip by the manufacturer\cite{Impinj}.
\begin{figure}
\centerline{\includegraphics[width=80mm]{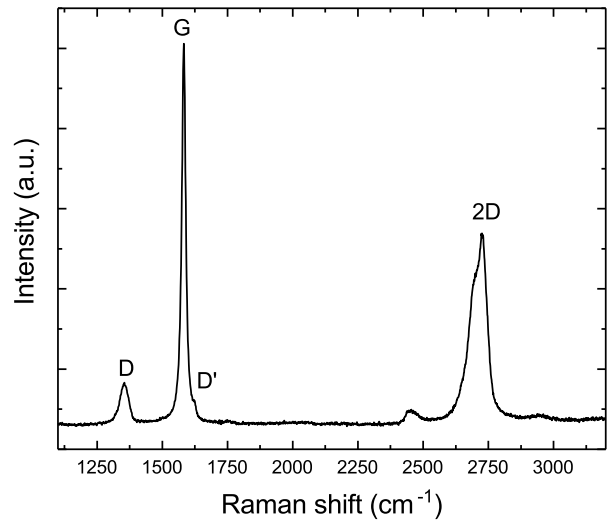}}
\caption{Representative Raman spectrum at 514nm for flakes processed for 70 cycles.}
\label{Fig:Raman}
\end{figure}
\begin{figure}
\includegraphics[width=80mm]{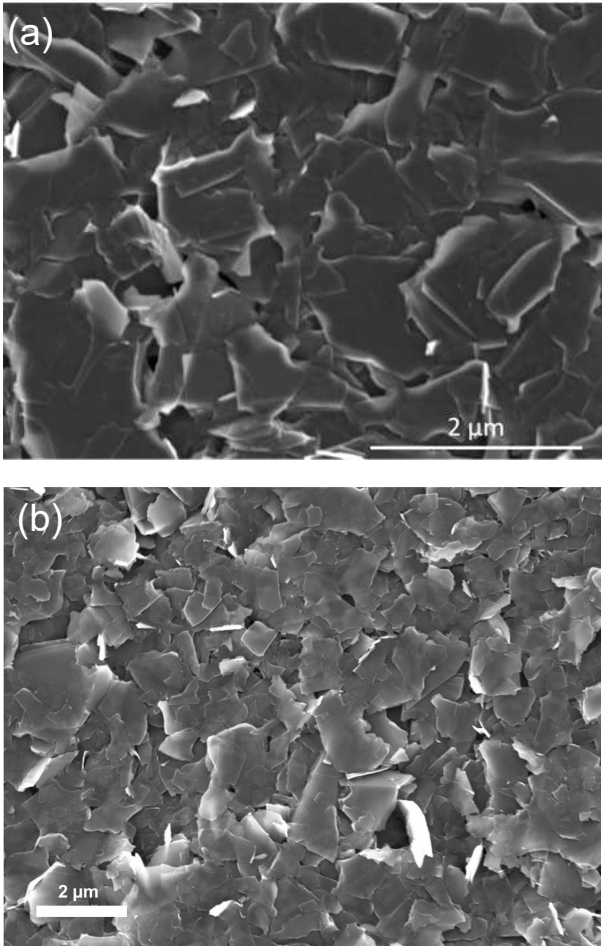}
\caption{SEM images of a)SP film on Kapton; b) SC film on paper}
\label{Fig:SEM}
\end{figure}
\begin{table}
\centering
\begin{tabular}{ | m{3cm} | m{2cm}| m{0.7cm} | m{0.7cm} | m{0.7cm} | m{0.7cm} | }
\hline
Transponder& Z$_{ant}$ (Ohms) & L$_Z$ (dB) & $\eta$ (dB) & D$_{tag}$ (dB) & R$_{read}$ (m) \\
\hline
Antenna with FLG inductive loop& 77.5 + j138 & -2.3 & -5.4 & 3.2 & 8.9 \\
\hline
Hybrid antenna& 17.2 + j136 & -0.0 & -4.0 & 3.0 & 13.1 \\
\hline
\end{tabular}
\caption{Simulated parameters of the two tag antennas in Figs.\ref{Figure1},\ref{Figure2} at 915MHz}
\label{Table 1}
\end{table}

Table I indicates that the transponder with a hybrid antenna has a longer read range (13.1m). This is due to both better impedance match between antenna and microchip, and higher radiation efficiency.
\begin{figure}
\centerline{\includegraphics[width=90mm]{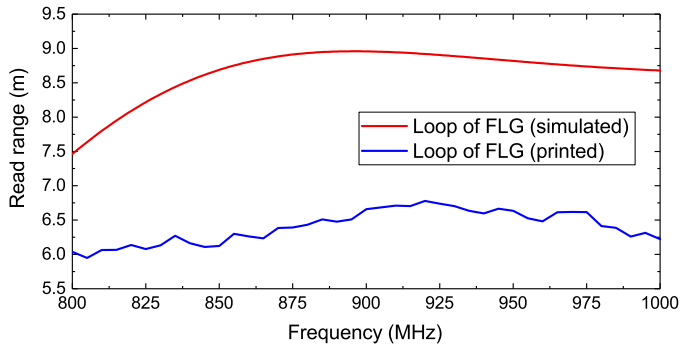}}
\caption{Simulated and measured read range as a function of frequency for antenna with FLG inductive loop}
\label{range_loop}
\end{figure}
\begin{figure}
\centerline{\includegraphics[width=90mm]{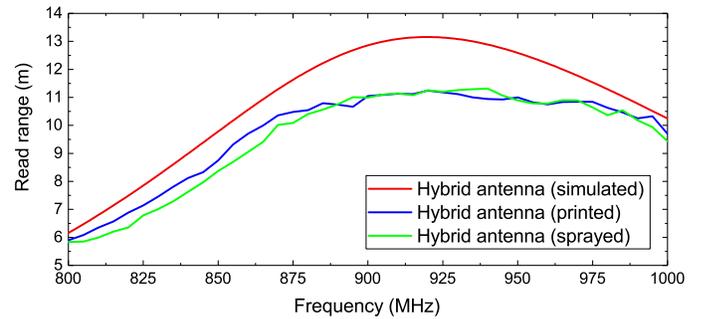}}
\caption{Simulated and measured read range as a function of frequency for hybrid antenna}
\label{range_Al}
\end{figure}
\begin{figure*}
\centerline{\includegraphics[width=150mm]{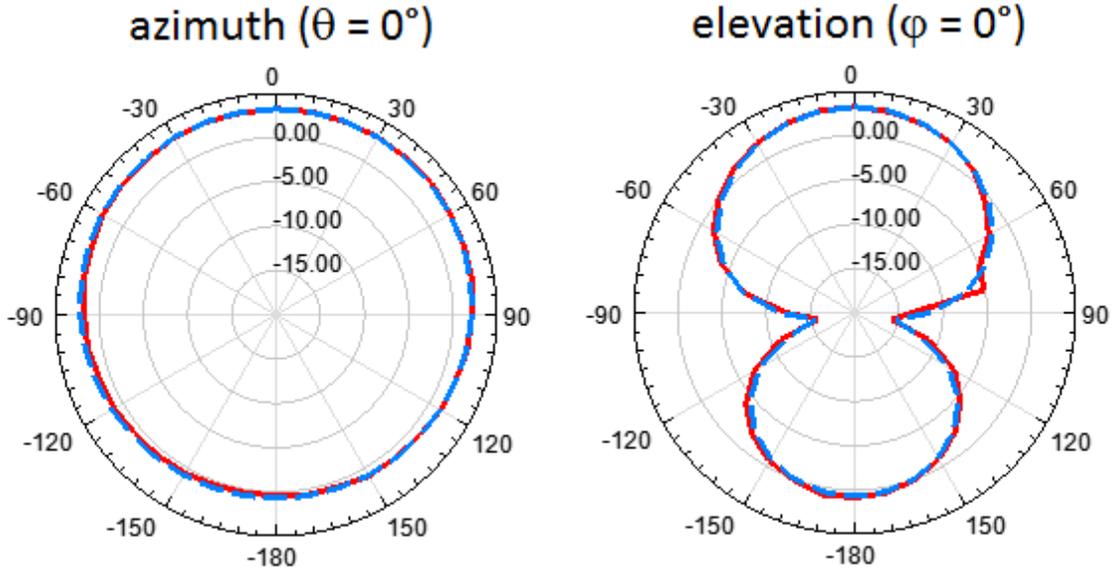}}
\caption{Measured (solid line) and simulated (dashed line) directivity of antenna with FLG inductive loop in azimuth and elevation plane}
\label{directivity_loop}
\end{figure*}
\begin{figure*}
\centerline{\includegraphics[width=150mm]{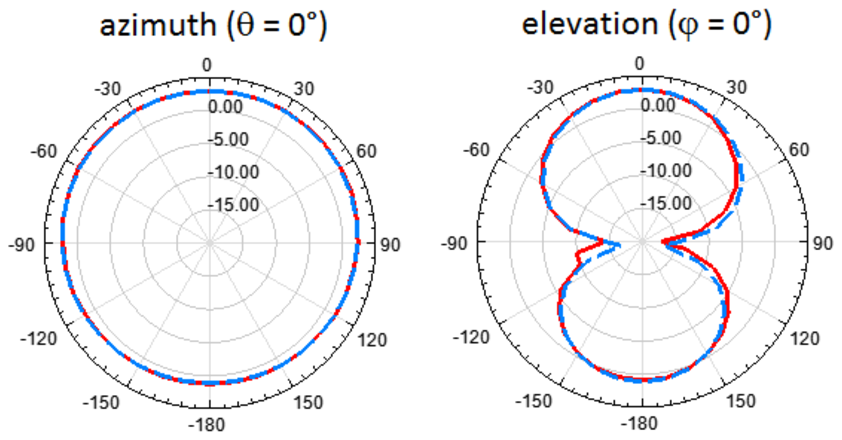}}
\caption{Measured (solid line) and simulated (dashed line) directivity of hybrid antenna in azimuth and elevation plane}
\label{directivity_Al}
\end{figure*}

Based on the design optimized by simulations, FLG antennas are fabricated either by screen printing or spray coating. Two inks suitable for screen printing and spray coating are formulated by adding different amounts of rheology modifiers after exfoliation of graphite to tune the ink viscosity. Graphite flakes (Timrex KS25) are added to deionized (DI) water at a concentration$\sim$100g/L and sodium deoxycholate ($\sim$5g/L). The mixture is processed using a microfluidizer (M-110P) at 207MPa for 70 cycles. One cycle is defined as one pass of the liquid mixture through the interaction chamber, where high shear rate ($\sim$10$^8$s$^{-1}$) is applied\cite{Karagiannidis2017}. The exfoliated graphite flakes have a lateral size distribution peaked at$\sim$1$\mu m$ and thickness$\sim$12nm\cite{Karagiannidis2017}. Fig.\ref{Fig:Raman} plots a representative Raman spectrum, acquired by a Renishaw inVia at 514nm excitation, of the processed material after microfluidization. The 2D peak consists of two components (2D$_2$, 2D$_1$). Their intensity ratio I(2D$_2$)/I(2D$_1$), changes from$\sim$1.5 for the starting graphite to$\sim$1.2, indicating exfoliation, but not complete to SLG\cite{Karagiannidis2017,FerrariPRL}. Following microfluidization, carboxymethylcellulose (CMC) sodium salt is added at a concentration$\sim$10g/L to prepare a screen printable (SP) ink and$\sim$5g/L for the spray coating (SC) one. CMC acts as rheology modifier imparting to the SP-ink a viscosity ranging from$\sim$570mPa s at 100s$^{-1}$ to$\sim$140mPa s at 1000s$^{-1}$, and to the SC one$\sim$220mPA s at 100s$^{-1}$ to 60mPa s at 1000s$^{-1}$.

The SP-ink is used to form FLG films both for antennas with FLG inductor and hybrid antennas on Kapton using a screen printer (Kippax KPX-2012) equipped with a 90 mesh per inch screen. These are then annealed at 265$^{\circ}$C for 10 minutes to remove the binder and increase conductivity. R$_S$ of the printed antennas measured using a four-point probe is$\sim$5$\Omega/ \square$, reduced to$\sim$3$\Omega/ \square$ after annealing at 265$^{\circ}$C for 10mins. Fig.\ref{Fig:SEM}a is a scanning electron microscope (SEM) image of the printed film after annealing. Annealing at higher temperatures or for longer times further reduces R$_S$, however it causes delamination from Kapton, making the antenna not usable.

The SC ink is used for hybrid antennas and sprayed onto 3 substrates: 1) Polyethylene Naphthalate (PEN), Q65HA-125$\mu$m); 2) multicoated matt art paper (Lumisilk-120$\mu$m); 3) uncoated printing paper (Tesorp). The substrate is cut into the shape of the simulated antenna. SC is performed using a hand held manual spray pen for$\sim$5s, while moving over the antenna area, so that ink covers the whole substrate, resulting in a self-standing antenna. Air pressure is kept constant and the spraying distance is$\sim$20cm. The dry thickness of one pass is$\sim$15-18$\mu$m. A SEM image of a FLG film on paper is in Fig.\ref{Fig:SEM}b.

The uncoated paper completely absorbs the water from the ink and the samples are dried and flattened using a hot press at$\sim$130 $^{\circ}$C. The samples are then calendered using a cylinder press with one steel roller and one hard rubber roller, generating a pressure$\sim$80bar ($\sim$36kN/m). The compression is performed at 2m/min and up to 3 times. The adhesion of the dry ink on plastic and multicoated paper is not optimal, so this process is only done for uncoated printing paper where the ink is more easily absorbed deep into the substrate. R$_S$ is measured by four-probe close to the centre of the antenna, where the highest conductivity is required, as shown in the simulations in Figs.\ref{Figure1},\ref{Figure2}. R$_S$ saturates at$\sim$3.6$\Omega/ \square$ after 2 spray passes. Further calendaring or additional coating do not further reduce R$_S$. The reason is that paper fibres limit the conducting pathways available for the FLG flakes as the ink is absorbed into the substrate before it can dry, due to the FLG concentration and the evaporation of water. SEM images of SP films on Kapton and SC on paper are shown in Figs.\ref{Figure1}a,b.

For the transponder with FLG inductor, the microchip is glued directly to the antenna using Ag paste. For the hybrid system, the Al inductive loop is fabricated similarly to conventional dipole transponders\cite{FinkBook}, i.e. by etching Al on polyethylene terephthalate (PET)\cite{FinkBook}. The microchip is subsequently attached onto the Al loop using anisotropic conducting adhesive (ACA)\cite{LinJMS43} and the loop is attached on the antenna with adhesive tape.

Using an Al inductor loop not only improves impedance matching in terms of conjugate impedance, but also reduces signal attenuation between antenna and microchip. Indeed, forming a contact between FLG antennas and microchip is challenging, especially considering the small ($\sim$400$\mu$m$\times$250$\mu$m) contact pads of an RFID microchip. ACA, typically used with metallic tags\cite{LinJMS43}, does not necessarily work on FLG, due to the temperature and pressure required by the bonding process\cite{LiuCONF}. Therefore, similar to Ref.\cite{HuanSR5}, for the  antenna wit FLG inductive loop we use Ag paint to establish an electrical contact between FLG films and RFID chip, Fig.\ref{Figure1}a. Conversely, in our hybrid design, the printed FLG antenna and the RFID chip are connected through the Al loop and no bonding or Ag paint is required between loop and FLG antenna. Therefore, conventional ACA can be used to bond the RFID chip to the Al loop.

Fig.\ref{Figure1} shows image and simulated current distribution of the antenna with parallel inductor implemented as an opening on the FLG conductor. The current is concentrated around the opening or the loop inductor of the transponder, Fig.\ref{Figure1}b. The hybrid antenna is shown in Fig.\ref{Figure2}a. Fig.\ref{Figure2}b is the corresponding simulated current distribution. The highest density of current is in the metal conductor, thus maximizing power transfer to the microchip, therefore improving the reading range. The antennas are measured with a Tagformance$^{\text{TM}}$ UHF RFID measurement system\cite{Tagformance} in its anechoic cabinet. The evaluation is based on measuring the activation level of the transponder in a fixed and known setup\cite{Niki_online,Pursula2007}. The transponders are attached on a piece of Styrofoam, acting as radiation-transparent support. The measured activation level is then used to calculate the theoretical reading range (i.e. the maximum range) in Figs.\ref{range_loop},\ref{range_Al}. The simulated reading ranges are also included for comparison.

For all antennas, the measured read range is shorter than simulations. However, for the hybrid antenna the discrepancy is smaller. A possible cause for this is the roughness of the edges in the SP antennas. Fig.\ref{Figure1} indicates that the current concentrates on the edges of the opening or the loop in the middle. Thus any added resistivity there has a significant impact on losses. This also explains why the difference between simulations and measurements is greater for antennas with FLG inductive loop. These also use Ag paste as the conductor between antenna and microchip. The connections between Ag paste and FLG, as well as between Ag paste and microchip contact pads, are likely to introduce additional contact resistance, hence signal attenuation.

Fig.\ref{range_Al} shows that the reading range of SP and SC antennas are almost identical. Only below$\sim$880MHz the distance of SP antennas is$\sim$10\% smaller than SC, showing how both deposition methods are suitable for the realization of FLG antennas.

The radiation patterns are also measured with the Tagformance$^{\text{TM}}$ system. Figs.\ref{directivity_loop},\ref{directivity_Al} plot the measured directivities (solid red lines) as compared to simulations (dashed blue lines). As the absolute directivity is difficult to measure, the measured radiation patters are normalized to the simulated ones at $\phi$=0, $\theta$=0.

Radiation patterns, both simulated and measured, reveal a small difference compared to an ideal dipole antenna. The radiation pattern is not perfectly round on the azimuth plane. The difference in directivity between 0 and 180$^o$ is 2.8dB for the FLG inductive loop antenna and 1.7dB for the hybrid one. This can be seen also on the elevation plane and as the maximum directivity values in Table 1 being above the theoretical one of a dipole antenna, 2.15dBi (decibels relative to isotropic radiator)\cite{PozaBOOK}. This can be attributed to the asymmetry of the transponders combined with the FLG R$_S$.
\section{Conclusions}
UHF RFID transponders with screen-printed and sprayed FLG antennas were designed, fabricated and tested. Read ranges$\sim$6.7 and 11.1m were measured for antennas with FLG inductive loop and hybrid antennas, respectively. The transponders operate at the frequency bands reserved for UHF RFID: 865.6-867.6 MHz (Europe) and 902-928 MHz (USA, Japan). The hybrid antenna has reading performance superior to previously reported graphene-based RFID tags\cite{Akbari2016, Arapov2016, Kopyt2016,PamNC9} and  comparable with commercial ones\cite{Datasheet}. It also avoids the need for a direct contact between FLG film and microchip, making the fabrication of FLG antennas compatible with existing industrial processes.
\section{Acknowledgements}
We acknowledge funding from EU Graphene Flagship, the French RENATECH network, ERC Grant Hetero2D, EPSRC Grants EP/K01711X/1, EP/K017144/1, EP/N010345/1, and EP/L016087/1.

\end{document}